\documentclass[12pt,preprint]{aastex}
\slugcomment{submitted to ApJ Letters}

\shorttitle{Companion Stars of SNe Ia}
\shortauthors{Han}

\begin{document}
\title{Companion Stars of Type Ia Supernovae}


\author{Z. Han}
\affil{National Astronomical Observatories / Yunnan Observatory,
       the Chinese Academy of Sciences, Kunming, 650011, China}
\email{zhanwenhan@hotmail.com}

\begin{abstract}
The WD+MS channel of the single-degenerate scenario is currently
favourable for progenitors of type Ia supernovae (SNe Ia).
Incorporating the results of detailed binary evolution calculations
for this channel into the latest version of a binary population 
synthesis code, I obtained the distributions of many properties 
of the companion stars at the moment of SN explosion. The properties can
be verified by future observations.
\end{abstract}


\section{Introduction}

Type Ia supernovae (SNe Ia) have been 
 used as a {\it calibrated} candle to probe the dynamics of the universe, 
leading to a significant progress in cosmology, i.e. the determination of
$\Lambda$ and $\Omega$ \citep{rie98,per99}.
They are believed to be thermonuclear explosions of carbon-oxygen (CO)
white dwarfs (WDs). However the nature of their progenitors 
has remained unclear, and this still raises doubts as to the calibration
which is purely empirical and based on the nearby SN Ia sample.
Based on the characteristics of observed SNe Ia (e.g. light curves,
chemical stratification), it seems most likely that they occur
when the accreting CO WDs reach the Chandrasekhar limit, and
sub-Chandrasekhar models appear not to be consistent with the observations
of SNe Ia. Chandrasehar-mass WDs can be created through the 
single-degenerate scenario, where the CO WD accretes mass from
a non-degenerate companion \citep{nom84,hac99}, or the double degenerate
scenario, where two CO WDs with a total mass larger than the Chandrasekhar
mass coalesce \citep{ibe84,web87}\footnote{Note, it is quite likely that
the merger product experiences core collapse rather than a thermonuclear
explosion.}. Recent observations indicate that there is
a wide spread of delay time between star formation and SN Ia explosion, and
this imply that there exist two populations of progenitors, a ``prompt''
one with a delay time less than $\sim 0.1\,{\rm Gyr}$ 
and a ``tardy'' one with a delay time of $\sim 3\,{\rm Gyr}$ \citep{man06}. 

Using the archival data prior to explosion from Hubble Space Telescope (HST)
and {\em Chandra}, \citet{vos08} tried to search for a 
possible progenitor of SN2007on (type Ia) in elliptical galaxy NGC 1404.
They discovered an object at the position of SN2007on. The X-ray
luminosity of the object and the non-detection in the optical images
are fully consistent with the single-degenerate scenario, and the
the discovery therefore favours the single-degenerate scenario.  
However, new {\em Chandra} X-ray observations \citep{roe08} 
and detailed astrometry of the site of SN2007on showed that
there appears to be an offset between the SN and the X-ray source,
and the offset means a small probability ($\sim 1\%$) 
of the X-ray source being related to the SN.
However, the X-ray source is not unconnected with the SN, as 
the X-ray source has dimmed after the explosion
and the source before the explosion showed an excess of soft X-ray photons
relative to the other sources in the field. 
Kepler's 1604 supernova is suggested to be of type Ia.
\citet{rey07} made a deep {\em Chandra} observation of Kepler's supernova
remnant and concluded that Kepler's SN is a SN Ia with circumstellar
medium (CSM) interaction. This might indicate that the progenitor is
a massive (young) star. \cite{mao08} searched
HST pre-explosion images for NGC 1316, 
a radio galaxy in the Fornax cluster
and a prolific producer of SNe Ia, for evolved intermediate-mass progenitor
stars. The pre-explosion images (3 years before explosion) of the sites
of SN2006dd (a normal SN Ia) and SN2006mr (likely a subluminous SN Ia)
show no potential luminous stellar progenitors. 
More effort is still needed in identifying the progenitors of SNe Ia.

Among various possible progenitor models, 
the WD+MS channel of the single-degenerate scenario
is the most widely accepted one, in which
a CO WD accretes mass from its main-sequence (MS) star or
a slightly evolved subgiant in a close binary
system until it reaches a mass of $\sim 1.378M_\odot$ and 
explodes as a SN Ia \citep{nom84,li97,hac99,lan00,han04}.
The companion star should survive the explosion and show distinguishing 
properties, and it is therefore a promising method to test progenitor models
by identifying the surviving companion stars of SNe Ia.   

Tycho Brahe's 1572 supernova is a Galactic SN Ia. 
\citet{rui04} found in the remnant region that
Tycho G, a star similar to the Sun but with a lower gravity,
moves at more than three times the mean velocity 
of the stars there.
They argued that Tycho G could be the surviving companion of the 
supernova\footnote{However, 
\citet{fuh05} argued that there exists a possibility of 
Tycho G being a thick disk star coincidentally 
passing in the vicinity of the remnant of SN 1572.}.
Indeed, a surviving companion would have a high space velocity and  
evolve to a WD finally, and the 
single-degenerate scenario could potentially explain
the properties of halo WDs observed by \citet{opp01}, e.g.
their space density and ages \citep{hansen03,ber03}.
Furthermore, \citet{jus08} argued 
that the ultra-cool WDs observed by \citet{wol05} might have
been formed via the single-degenerate scenario, i.e. they might be
the remnants of the non-degenerate donor stars.
Note, however, there have been no conclusive proof yet 
that any individual object is the surviving companion of a SN Ia.

\citet{lan00} did detailed binary evolution
calculations for part of the WD+MS channel of 
the single-degenerate scenario, and presented
the properties, e.g. orbital velocities, luminosities, effective temperatures,
of the companion stars at the moment of SN Ia explosion for the sample
WD binaries they studied, in which the Roche lobe overflow (RLOF) 
starts when the companion star
is in MS. However, the distributions of the properties have not been obtained.
\citet{can01} run a Monte Carlo scenario code to calculate the 
distributions of masses, luminosities and velocities of the companion
stars of SNe Ia. However, they have not done detailed binary 
evolution calculations for the production of SNe Ia, and this
results in big uncertainties \citep[see][]{han04}, e.g. the orbital 
velocities obtained for the companion stars from the WD+MS channel
are over 450 km/s, which is far too high. 

\citet[][hereafter HP04]{han04} carried out detailed binary 
evolution calculations
for the WD+MS channel for about 2300 close WD binaries, in which
RLOF starts when the companion star is in MS or in Hertzsprung gap
(i.e. slightly evolved). The study is comprehensive, and various
properties of the companion stars were obtained but not sorted for
publishing. In this {\em Letter}, I extract the properties from the data files 
of the calculations and incorporate them
into the latest version of the binary population synthesis (BPS) code 
developed for the study of various binary-related objects 
\citep{han95b,han95a,han98,han02,han03}, including the progenitors of
SNe Ia \citep{han04,han06}, and obtain the distributions of the
properties.

\section{The distributions of properties of the companion stars}

In the single-degenerate scenario, the progenitor of a SN Ia is a close
WD binary system, which has most likely emerged from common envelope (CE)
evolution \citep{pac76} of a giant binary system. 
During the CE evolution, the envelope engulfs the core (here a CO WD)
of the giant and the secondary, and the orbital energy released in the
spiral-in process (i.e. orbital decay) is used to overcome the 
binding energy of the CE in order to eject it. For the CE evolution,
I have two parameters: $\alpha_{\rm CE}$ the
CE ejection efficiency, i.e.\ the fraction of the
released orbital energy used to overcome the binding energy, 
and $\alpha_{\rm th}$, which defines the fraction of the thermal
energy contributing to the binding energy of the CE.
As in previous studies, I adopted $\alpha_{\rm CE}=\alpha_{\rm th}=1.0$,
which gives good matches between theory and observations for
many binary-related objects
\footnote{
The prescription adopted here for the CE evolution is different from
the $\lambda$ prescription, but appears to be more physical. 
See \citet{han95b}, \citet{dew00} and \citet{pod03b} for details.
}.

To obtain the distributions of properties of companion stars 
at the moment of SN explosion,
I have performed a detailed Monte Carlo simulation with 
the latest version of the BPS code.
The code follows the evolution of binaries
with their properties being recorded at every step.
If a binary system evolves to a WD+MS system, and if the
system, at the beginning of the RLOF phase, is located
in the SN Ia production regions in the plane of
($\log P^{\rm i}$, $M_2^{\rm i}$) for its $M_{\rm WD}^{\rm i}$, 
where $P^{\rm i}$, $M_2^{\rm i}$ and $M_{\rm WD}^{\rm i}$
are, respectively, the orbital period, the secondary's mass and 
the WD's mass of the WD+MS system at the beginning of the RLOF
(see Fig. 3 of HP04), I assume that a SN Ia is
resulted, and the properties of the WD binary at the moment of SN explosion
are obtained by interpolation in the 3-dimensional grid 
($M_{\rm WD}^{\rm i}$, $M_2^{\rm i}$, $\log P^{\rm i}$) of 
the $\sim 2300$ close WD binaries calculated in HP04. 

In the simulation, I follow the evolution of 100 million sample binaries
according to grids of stellar models of metallicity $Z=0.02$ and the
evolution channels leading to SNe Ia as described in HP04.
I adopted the following input for the simulation
\citep[see][]{han95b}.  (1) The
star-formation rate (SFR) is taken to be constant over the last 15\,Gyr.
(2) The initial mass function (IMF) of \citet{mil79} is adopted. 
(3) The mass-ratio distribution is taken
to be constant. (4) The distribution of
separations is taken to be constant in $\log a$ for wide binaries, where $a$ is
the orbital separation. (5) The orbits are assumed to be circular.

The simulation gives current-epoch-snapshot distributions 
of many properties of companion stars at the moment of SN explosion, e.g.
the masses, the orbital periods, the orbital separations, 
the orbital velocities, the effective temperatures, the luminosities, 
the surface gravities, the surface abundances, the mass transfer rates, 
the mass loss rates of the optically thick stellar winds. 
The simulation also shows
the initial parameters of the
primordial binaries and the WD binaries that lead to SNe Ia. 
Figs~\ref{vm}-~\ref{nm} are selected distributions that may be helpful
to identify the progenitors of SNe Ia.

\section{Discussion}

\begin{figure}
\includegraphics[width=5.5cm,angle=270]{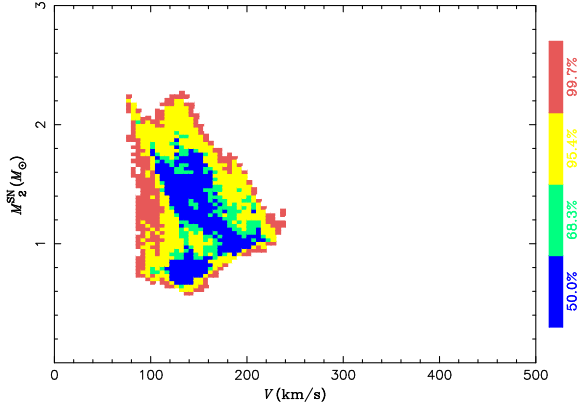}
\caption{A snapshot distribution of companion stars in the plane of 
($V$, $M_2^{\rm SN}$) at current epoch, where $V$ is the orbital velocity and
$M_2^{\rm SN}$ the mass at the moment of SN explosion. The number density
decreases from inner regions to outer regions. 
Regions, from inside to outside with corresponding
gradational grey or color in the legend (from bottom to top), 
together with the regions with higher number densities
contain 50.0\%, 68.3\%, 95.4\%, and 99.7\%
of all the systems, respectively.
\label{vm}}
\end{figure}

\begin{figure}
\includegraphics[width=5.5cm,angle=270]{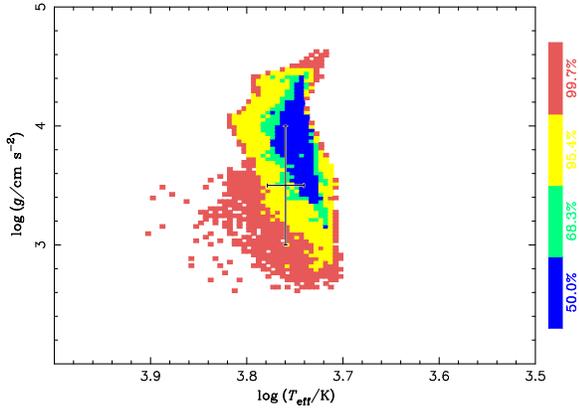}
\caption{Similar to Fig.~\ref{vm}, but in the plane of
($\log T_{\rm eff}$, $\log g$), where $T_{\rm eff}$ is the
effective temperature of companion stars at the moment  of SN explosion,
$\log g$ the surface gravity. The error bars denote the location
of Tycho G \citep{rui04}.
\label{tg}}
\end{figure}

Figs~\ref{vm} and ~\ref{tg} are the distributions of
the masses, the orbital velocities
\footnote{The Chandrasekhar-mass WD has an orbital velocity of
$\sim 50$ to $\sim 200\,{\rm km/s}$ for a corresponding 
companion star's mass of $\sim 0.6$ to $\sim 2.0M_\odot$ 
at the moment of SN explosion.}, 
the effective temperatures and
the surface gravities of companion stars at the moment of SN explosion. 
Tycho G was taken as the surviving companion star of 
Tycho Brahe's 1572 supernova by \citet{rui04}. It has a space velocity
of $136\,{\rm km/s}$, more than 3 times the mean velocity there.
Its surface gravity is $\log\, (g/{\rm cm}\, {\rm s}^{-2})=3.5\pm 0.5$,
while the effective temperature is $T_{\rm eff}=5750\pm 250 {\rm K}$.
The parameters are compatible with Figs~\ref{vm} and ~\ref{tg}.
As from our simulation, the recorded properties at each step
show that a primordial binary system
with a primary mass $M_{\rm 1i}\sim 4-5.5M_\odot$, a secondary mass
$M_{\rm 2i} \sim 2-3M_\odot$ and an orbital period 
$P_{\rm i} \sim 100-250\,{\rm d}$ would evolve to a close WD binary
system with a WD mass $M^{\rm i}_{\rm WD} \sim 0.8-1.2M_\odot$, 
a secondary mass $M^{\rm i}_{\rm 2} \sim 2-3M_\odot$, 
and an orbital period $P^{\rm i} \sim 1-4\,{\rm d}$.
The WD binary results in SN Ia explosion with companion parameters
($T_{\rm eff}$ and $\log g$ actually) in 
the range of Tycho G.

However, Figs~\ref{vm} and ~\ref{tg} are for that at the moment of
SN explosion, the distributions could be modified due to the explosion. 
\citet{mar00} presented several high-resolution two-dimensional numerical
simulations of the impacts of SN Ia explosion with companions.
The impact make the companion in the WD+MS channel
lose a mass of 0.15-0.17$M_\odot$, and receive a kick of
49 - 86${\rm km/s}$.
\citet{men07} adopted the simple analytic method of \citet{whe75}, 
and calculated the impact to survey the influence of the 
initial parameters of the progenitor's systems. 
With detailed stellar models and realistic 
separations that were obtained from binary evolution, 
they obtained an even lower 'stripped mass', 0.03-0.13$M_\odot$, but
a similar kick velocity 30-90${\rm km/s}$, which is perpendicular
to the orbital velocity.
A surviving companion star therefore has a mass lower 
by $\sim 0.1M_\odot$ and a space velocity larger by $\sim 10\%$
than that in Fig.~\ref{vm}.

The companion stars are out of thermal equilibrium at the moment of SN 
explosion. The equilibrium radii are typically larger by $\sim 50\%$ than
that at the moment of SN explosion. Therefore the surface gravity at
equilibrium should be lower than that in Fig.~\ref{tg}.
\citet{pod03a} systematically explored the evolution and appearance of a 
typical companion star that has been stripped and heated by the supernova
interaction during the post-impact re-equilibrium phase. Such a star
may be significantly overluminous or underluminous.
Fig.~\ref{tg} could be a starting point for further studies of this kind.

\begin{figure}
\includegraphics[width=5.5cm,angle=270]{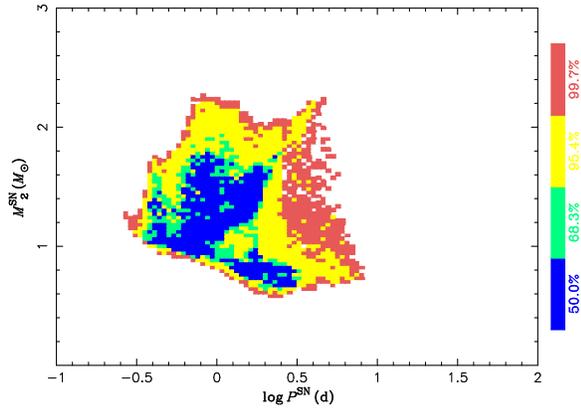}
\caption{Similar to Fig.~\ref{vm}, but in the plane of
($\log P^{\rm SN}$, $M_2^{\rm SN}$), where $P^{\rm SN}$ is the orbital 
period at the moment of SN explosion.
\label{pm}}
\end{figure}

\begin{figure}
\includegraphics[width=5.5cm,angle=270]{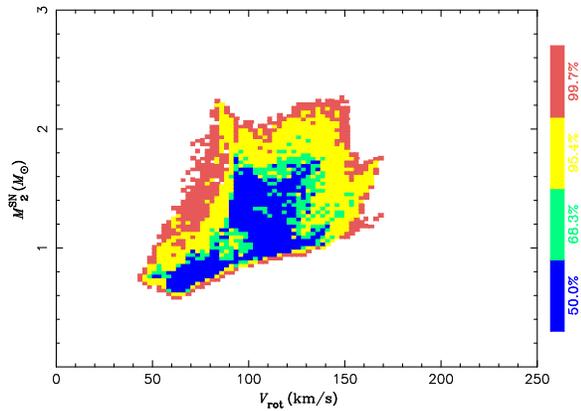}
\caption{Similar to Fig.~\ref{vm}, but in the plane of
($V_{\rm rot}$, $M_2^{\rm SN}$), where $V_{\rm rot}$
is the equatorial rotational velocity of companion stars 
at the moment of SN explosion.
\label{vrotm}}
\end{figure}

Fig.~\ref{pm} is the distribution of orbital periods of the WD+MS systems
at the moment of SN explosion. If I assume that the companion stars
co-rotate with their orbits, I obtain their distributions of equatorial 
rotational velocities (see Fig.~\ref{vrotm}). We see that
the surviving companion stars are fast rotators 
and their spectral lines should be broadened noticeably. 

\begin{figure}
\includegraphics[width=5.5cm,angle=270]{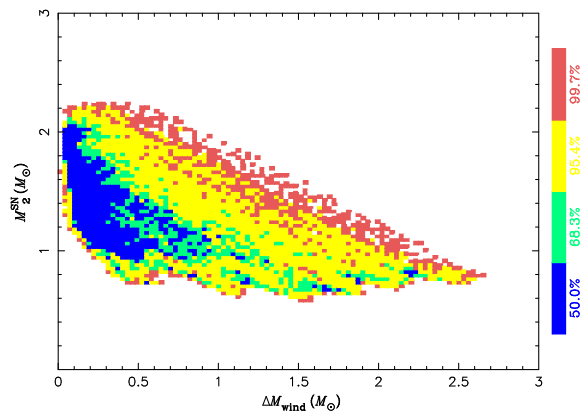}
\caption{Similar to Fig.~\ref{vm}, but in the plane of
($\Delta M_{\rm wind}$, $M_2^{\rm SN}$), where $\Delta M_{\rm wind}$
is the mass lost during the optically thick stellar wind phase
of the WD binaries.
\label{dmm}}
\end{figure}

Fig.~\ref{dmm} shows the distribution of masses lost during 
optically thick stellar wind phase. We see that a significant
amount of mass is lost in the wind. \citet{bad07} found that
the wind with a velocity above $200\,{\rm km/s}$, which
is believed to be reasonable, would excavate a large low-density
cavity around its progenitor. However, the fundamental properties
of seven young SN Ia remnants 
(Kepler, Tycho, SN 1006, 0509-67.5, 0519-69.0, N103B and SN 1885)
obtained by \citet{bad07} are incompatible with
such large cavities. A lower wind velocity or the consideration
of WD rotation could help to solve the problem
\citep{bad07}.

\begin{figure}
\includegraphics[width=5.5cm,angle=270]{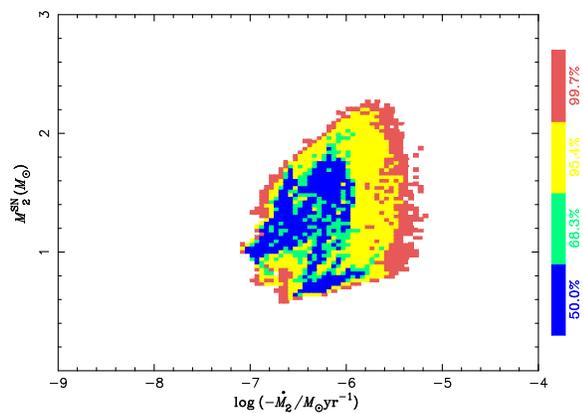}
\caption{Similar to Fig.~\ref{vm}, but in the plane of
($\log (-\dot M_2)$, $M_2^{\rm SN}$),  where $-\dot M_2$ is 
the mass transfer rate at the moment of SN explosion.
\label{dmdtm}}
\end{figure}

Fig.~\ref{dmdtm} presents the distribution of mass transfer rates
at the moment of SN explosion. The mass transfer rates can be converted
to X-ray luminosities of the systems
by $L_{\rm X}\sim \epsilon |\dot M |$, where 
$\epsilon=7\times 10^{18} {\rm erg/g}$ gives the approximate amount
of energy obtained per gram of hydrogen burnt into helium or 
carbon/oxygen. The luminosity of the X-ray source close to the
site of SN 2007on (4 years before the explosion) was
estimated to be $(3.3\pm 1.5)\times 10^{37}\, {\rm erg/s}$
\citep{vos08}, corresponding to a mass accretion rate of 
$\sim 10^{-7}\,M_\odot/{\rm yr}$, which is consistent with
Fig.~\ref{dmdtm}.  

QU Carinae, a cataclysmic variable (CV), is suspected
to be a SN Ia progenitor. It has a WD mass of $\sim 1.2M_\odot$,
a once-reported orbital period of $0.45\, {\rm d}$, and more importantly,
a very high mass transfer rate of $\sim 10^{-7}\,M_\odot/{\rm yr}$
\citep{kaf08}. 
These properties are consistent with Figs~\ref{pm} and ~\ref{dmdtm}.
However, BF Eridani, another CV with $M_{\rm WD}\sim 1.28M_\odot$,
$M_2\sim 0.52M_\odot$ and $P\sim 0.27\,{\rm d}$ \citep{neu08}, 
appears not to be a SN Ia progenitor.

\begin{figure}
\includegraphics[width=5.5cm,angle=270]{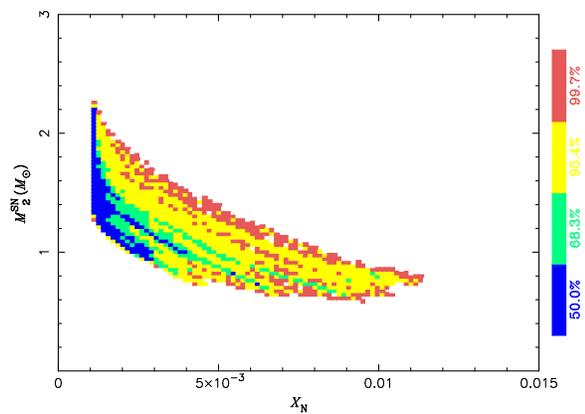}
\caption{Similar to Fig.~\ref{vm}, but in the plane of
($X_{\rm N}$, $M_2^{\rm SN}$), where $X_{\rm N}$ is the nitrogen mass
fraction at the surface of companion stars at the moment of SN explosion. 
\label{nm}}
\end{figure}

Fig.~\ref{nm} is the distribution of the surface nitrogen mass fraction
of companion stars at the moment of explosion. We see that nitrogen can be
significantly overabundant
(with corresponding underabundance of carbon due to the CN-cycle).
However, the surface can be seriously polluted by the ejecta of 
SN explosions.

The simulation in this {\em Letter} was made with
$\alpha_{\rm CE}=\alpha_{\rm th}=1.0$. If I adopt
a lower value for $\alpha_{\rm th}$, say, 0.1, the birth rate of SNe Ia
would be higher (a factor of 1.7) and the delay time from the star
formation to SN explosion is shorter.
This is because that binaries 
resulted from CE ejections tend to have shorter orbital periods for
a small $\alpha_{\rm th}$ and are more likely to locate in the SN Ia production
region (see Fig.3 of HP04). 
The companion stars with orbital velocity $V<110\,{\rm km/s}$ 
would be absent from Fig.~\ref{vm},
and the ones with $\log (g/{\rm cm\ s^{-2}}) <3.1$ would be
absent from Fig.~\ref{tg}.
This is due to that WD binaries with
long orbital periods are absent due to a small $\alpha_{\rm th}$.

The distributions are snapshots at current epoch for a constant SFR.
For a single star burst, most of the 
SN explosions occur between 0.1 and 1\,Gyr after the burst
(see Fig. 7 of HP04). The evolution of progenitor properties 
with time can be understood via Fig. 3 of HP04.
A delay time from 0.1 to 1\,Gyr corresponds to $M_2^{\rm i}$
of $\sim 3.2M_\odot$ to $\sim 1.8M_\odot$, and to
$M_{\rm WD}^{\rm i}$ of $\sim 1.2M_\odot$ to $\sim 0.67M_\odot$
for the WD+MS system, respectively.
As seen from Fig. 3 of HP04, the range of the orbital periods becomes
narrower from a delay time of 0.1 to 1.0\,Gyr. 
Those WD binaries result in SN Ia explosions via RLOF.
Consequently, the range of progenitor properties at the moment of SN Ia, e.g.
the orbital velocities of companion stars, the surface gravities, 
the equatorial rotational velocities, the mass transfer rates, becomes
narrower with time. The mass transfer rate would be smaller with time
as $M^{\rm i}_2$ becomes smaller.

\acknowledgments
I thank an anonymous referee for his/her comments 
which help to improve the paper.
I thank Ph. Podsiadlowski for stimulating discussions.
This work was in part supported by the Natural Science Foundation of China 
under Grant Nos 10433030, 10521001 and 2007CB815406.

\clearpage

\begin{thebibliography}{}
\bibitem[Badenes et al.(2007)]{bad07}
 Badenes C., Hughes J.P., Bravo E., Langer N., 2007, \apj, 662, 472
\bibitem[Bergeron(2003)]{ber03}
 Bergeron P., \apj, 586, 201
\bibitem[Canal, M\'endez \& Ruiz-Lapuente(2001)]{can01}
 Canal R., M\'endez J., Ruiz-Lapuente R., \apjl, 550, L53
\bibitem[Dewi \& Tauris(2000)]{dew00}
 Dewi J.D.M., Tauris T.M., 2000, \aap, 360, 1043
\bibitem[Fuhrmann(2005)]{fuh05}
 Fuhrmann K., 2005, \mnras, 359, L35
\bibitem[Hachisu, Kato \& Nomoto(1999)]{hac99}
 Hachisu I., Kato M., Nomoto K., 1999, \apj, 522, 487
\bibitem[Han(1998)]{han98}
 Han Z., 1998, \mnras, 296, 1019
\bibitem[Han et al.(1995)]{han95a}
 Han Z., Eggleton P.P., Podsiadlowski Ph., Tout C.A., 
 1995, \mnras, 277, 1443
\bibitem[Han, Podsiadlowski \& Eggleton(1995)]{han95b}
 Han Z., Podsiadlowski Ph., Eggleton P.P., 1995, \mnras, 272, 800
\bibitem[Han et al.(2002)]{han02}
 Han Z., Podsiadlowski Ph., Maxted P.F.L., Marsh T.R., Ivanova N.,
 2002, \mnras, 336, 449
\bibitem[Han et al.(2003)]{han03}
 Han Z., Podsiadlowski Ph., Maxted P.F.L., Marsh T.R.,
 2003, \mnras, 341, 669
\bibitem[Han \& Podsiadlowski(2004)]{han04}
 Han Z., Podsiadlowski Ph., 2004, \mnras, 350, 1301 (HP04)
\bibitem[Han \& Podsiadlowski(2006)]{han06}
 Han Z., Podsiadlowski Ph., 2006, \mnras, 368, 1095
\bibitem[Hansen(2003)]{hansen03}
 Hansen B.M.S., \apj, 582, 915
\bibitem[Iben \& Tutukov(1984)]{ibe84}
 Iben I.Jr., Tutukov A.V., 1984, \apjs, 54, 335
\bibitem[Justham et al.(2008)]{jus08}
 Justham S., Wolf C., Podsiadlowski Ph., Han Z., 
 2008, \aap, submitted
\bibitem[Kafka, Anderson \& Honeycutt(2008)]{kaf08}
 Kafka S., Anderson R., Honeycutt R.K., 2008, \apj, submitted
 (astro-ph/0801.3638)
\bibitem[Langer et al.(2000)]{lan00}
 Langer N., Deutschmann A., Wellstein S., H\"{o}flich P., 2000, 
 \aap, 362, 1046
\bibitem[Li \& van den Heuvel(1997)]{li97}
 Li X.D., van den Heuvel E.P.J., 1997, \aap, 322, L9
\bibitem[Mannucci, Della Valle \& Panagia(2006)]{man06}
 Mannucci F., Della Valle M., Panagia N., 2006, \mnras, 370, 773
\bibitem[Marietta, Burrows \& Fryxell(2000)]{mar00}
 Marietta E., Burrows A., Fryxell B., 2000, \apjs, 128, 615
\bibitem[Meng, Chen \& Han(2007)]{men07}
 Meng X., Chen X., Han Z., 2007, \pasj, 59, 835
\bibitem[Maoz \& Mannucci(2008)]{mao08}
 Maoz D., Mannucci F., 2008, MNRAS, submitted (astro-ph/0801.2898)
\bibitem[Miller \& Scalo(1979)]{mil79}
 Miller G.E., Scalo J.M., 1979, \apjs, 41, 513
\bibitem[Neustroev \& Zharikov(2008)]{neu08}
 Neustroev V.V., Zharikov S., 2008, MNRAS, submitted
 (astro-ph/0801.1082)
\bibitem[Nomoto, Thielemann \& Yokoi(1984)]{nom84}
 Nomoto K., Thielemann F.,Yokoi K., 1984, \apj, 286, 644
\bibitem[Oppenheimer et al.(2001)]{opp01}
 Oppenheimer B.R., Hambly N.C., Digby A.P., Hodgkin S.T., Saumon D.,
 2001, Science, 292, 698
\bibitem[Paczy\'nski(1976)]{pac76} 
 Paczy\'nski B., 1976, in Eggleton P.P., Mitton S., Whelan J.,
 eds, Structure and Evolution of Close Binaries. Kluwer, Dordrecht, p. 75
\bibitem[Perlmutter et al.(1999)]{per99}
 Perlmutter S. et al., 1999, \apj, 517, 565
\bibitem[Podsiadlowski(2003)]{pod03a}
 Podsiadlowski Ph., 2003, ArXiv:astro-ph/0303660
\bibitem[Podsiadlowski, Rappaport \& Han(2003)]{pod03b}
 Podsiadlowski Ph., Rappaport S., Han Z., 2003, \mnras, 341, 385
\bibitem[Reynolds et al.(2007)]{rey07}
 Reynolds S.P., et al., 2007, \apjl, 668, L135
\bibitem[Roelofs et al.(2008)]{roe08}
 Roelofs G., Bassa C., Voss R., Nelemans G., 2008, MNRAS, submitted 
 (astro-ph/0802.2097)
\bibitem[Riess et al.(1998)]{rie98}
 Riess A. et al., 1998, \aj, 116, 1009
\bibitem[Ruiz-Lapuente et al.(2004)]{rui04}
 Ruiz-Lapuente P., et al., \nat, 431, 1069
\bibitem[Voss \& Nelemans(2008)]{vos08}
 Voss R., Nelemans G., 2008, \nat, 451, 802
\bibitem[Webbink \& Iben(1987)]{web87}
 Webbink R.F., Iben I.Jr., 1987, in Philipp A.G.D., Hayes D.S., Liebert J.W.,
    eds, IAU Colloq. No. 95, Second Conference on Faint Blue Stars.
    Davis Press, Schenectady, p.445
\bibitem[Wheeler, Lecar \& McKee(1975)]{whe75}
 Wheeler J.C., Lecar M., McKee C.F., 1975, \apj, 200, 145
\bibitem[Wolf(2005)]{wol05}
 Wolf C., 2005, \aap, 444, L49
\end{thebibliography}
\end{document}